\input harvmac

\Title{}{A Neutrino Mass Matrix with Vanishing $\mu$-$\mu$ and
$\tau$-$\tau$ Entries}
\centerline{Sheldon Lee Glashow\footnote{$^\dagger$}{Electronic
address: {\tt slg@bu.edu}}}

\bigskip\centerline{Department of Physics}
\centerline{Boston University}\centerline{Boston, MA 02215, USA}
\vskip .4 in

We revisit  our earlier proposal for the form of the neutrino mass
matrix: a two-zero ansatz wherein the CP-violating PMNS phase $\delta$
plays a surprisingly important role. We review its observable
consequences and show how our ansatz follows from a softly-broken
symmetry (muon number minus tau lepton number) in a see-saw model with
three Higgs doublets.

\Date{10/07}

\nref\rb{P.H. Frampton, S.L. Glashow, D. Marfatia, Phys.Lett.B536:79,2002.}
\nref\ra{A. Zee, Phys.Lett.B93:389,1980.}
\nref\rc{W. Grimus, S. Kaneko, L. Lavoura, JHEP 0601:110,2006,
 {\tt hep-ph/0510326.}}
\nref\rd{G.L. Fogli {\it et al.,} Nucl.Phys.Proc.Suppl.168:341,2007.}
\nref\re{H.V.  Klapdor-Kleingrothaus, Phys.Lett.B586:198,2004.}
\nref\rg{{\it E.g.,} J. Bonn {\it et al.,} {\tt hep-ph/0704.3930}.}
 
We examine neutrino masses and mixings in a see-saw scenario where the
neutrino mass matrix has a specific and predictive form: case $C$ of
reference\ \rb, one of several two-zero textures we had discusses
earlier.  Our model involves the standard model leptons, three neutral
singlet states to enable the see-saw mechanism and three Higgs
doublets.  A softly-broken flavor symmetry (muon number minus tau
lepton number) is imposed on both the Yukawa couplings of the Higgs
doublets and the large Majorana bare masses of the singlet neutrinos.
Flavor quantum numbers of the Higgs' and heavy neutrinos are
judiciously assigned so that Higgs vev's generate flavor-conserving
charged lepton masses and, via a see-saw, a neutrino mass matrix of
the desired form. Departures from this form are entirely negligible,
being of order $m/M$ where $m$ is the mass scale of light neutrinos
and $M$ that of the heavy singlets.  Our flavor symmetry is assumed to
be broken explicitly at dimension-2 by Higgs mass terms (as well as by
their vevs) so that no leptonic axion arises. After a brief
introduction, we recapitulate the observable implications of our
ansatz\rb\ and sketch the details of our model.

We employ a  flavor basis  wherein the mass matrices for charged leptons
and neutrinos  are:
 \eqn\ea{{\cal M}_\l=\pmatrix{m_e&0&0\cr 0&m_\mu&0\cr 0&0&m_\tau\cr}
\quad   {\cal M}_\nu= 
{\cal U}\pmatrix{m_1&0&0\cr 0&m_2&0\cr 0&0&m_3\cr}{\cal U}^{TR}}
with  $\cal U$  the conventionally defined PMNS matrix and $m_i$
 the (complex) neutrino masses.
It would be tedius to enumerate  the many attempts 
that have been made to find a simple and
viable texture for ${\cal M}_\nu$. I  mention just three:
$$\pmatrix{0&A&B\cr A&0&C\cr B&C&0\cr}\,,\quad
  \pmatrix{A&B&B\cr B&C&D\cr B&D&C\cr}\,,\quad
  \pmatrix{0&A&-A\cr A&B&0\cr -A&0&-B\cr}\,.$$
The first of these, the Zee ansatz\ra, 
 although  appealingly simple and readily realized, is disfavored by
  experiment. The second  is $\mu$-$\tau$ symmetric\rc\  and is
presently  both viable and faddish.
 Its special cases lead to `bimaximal' or
  `tribimaximal' neutrino mixing, but it implies
$\cos{2\theta_1}=0$ (hence maximal atmospheric oscillations) and
$\sin{\theta_2}=0$ (which implies no CP violation in oscillation
 phenomena), results which may not agree with future data.
 The third ansatz is  $\mu$-$\tau$
  antisymmetric. With two degenerate neutrino masses  it is patently
  unacceptable, but like the second ansatz 
it may serve as a sensible  starting point
about which to perturb\rc.
 
Our favored form of $\cal M$ is  a modified Zee ansatz with a
non-vanishing $e$-$e$ entry:
\eqn\eb{ {\cal M}= \pmatrix{A&B&C\cr B&0&D\cr C&D&0\cr}.}
 In general, such a neutrino mass matrix is  CP-violating
because  phase redefinitions may make any three (but not all) of the
four complex parameters real. Thus, there are four relations among the
nine physically meaningful parameters (three masses, four PMNS
parameters and two Majorana phases). These relations are encapsulated
by  the two complex equations:
\eqna\ec
$$\eqalignno{(c_1s_3+ s_1c_3\hat{s}_2)^2\,m_1
+ (c_1c_3-s_1s_3\hat{s}_2)^2\,m_2 +s_1^2c_2^2\,m_3
&=0\,,&\ec a\cr
(s_1s_3-c_1c_3\hat{s}_2)^2\,m_1+ (s_1c_3+c_1s_3\hat{s}_1)^2\,m_2
+c_1^2c_2^2\,m_3 
&=0\,. &\ec b\cr}$$
Here, $s_i\equiv \sin{\theta_i}$ and $c_i\equiv \cos{\theta_i}$,
whereas $\hat{s}_2$ denotes $e^{i\delta}\sin{\theta_2}$. 
We  use the notation
$\theta_1\equiv \theta_{23}$ and its cyclic permutations.

We employ two priors to deduce the observable consequences 
of eqs.\ec{a,b}.  The
subdominant angle is known to be small:\foot{Here and elsewhere we use
experimental results quoted by Fogli {\it et al.}\rd, such as:\hfil\break
$\Delta_s\simeq 2.6\times10^{-3}\;\rm eV^2$; $\Delta_a\simeq
7.92\times 10^{-3}\;\rm eV^2$
and $\tan{(2\theta_3)}\simeq 2.5$.}\ 
$s_2^2 < 0.03$. Errors  incurred  by omitting quadratic terms
in $s_2$  are readily estimable and correspondingly
small.  With  this approximation
eqs.\ec{a}\ may be recast as:
\eqna\ed
$$\eqalignno{s_3^2\,m_1+c_3^3\,m_2 + m_3 &=0\,, &\ed a\cr
 \sin{(2\theta_3)}\,\left(m_1-m_2\over 2m_3\right)\,
\hat{s}_2&=\cot{(2\theta_1)} \,. &\ed b\cr}$$
An immediate consequence of eq.\ed{a}\  is an obligatory  inverted
neutrino mass hierarchy. It is equivalent to the relation:
\eqn\ee{ m_1=-m_3+ac_3^2 e^{-i\delta}\quad{\rm and}\quad 
m_2=-m_3-as_3^2e^{-i\delta}\,,}
with $a$ real and positive and $\delta$ arbitrary, but soon to be
identified with the PMNS phase.

Our  second prior is the relative smallness of the solar
squared-mass parameter: $\Delta_s\simeq 0.03\,\Delta_a$.
While we could easily retain  small terms involving  $\Delta_s$
so as to  accomodate its observed value,  they
 would needlessly complicate our calculations without substantially
altering  our results. Thus we put
$|m_2|=|m_1|$ (hence $\Delta_s=0$) to  obtain:
\eqn\ef{ a= {2m_3\cos{\delta}\over \cos{(2\theta_3)}}\quad{\rm with}\quad
\delta /2\ge \phi\ge -\pi/2\, ,}
and
\eqn\eg{ |m_1|^2=|m_2|^2= m_3^2 (1+\tan^2{(2\theta_3)}\cos^2{\delta})\,.}
With these results, we  rewrite eq.\ed{b}\  as:
\eqn\esnd{\cot{(2\theta_1)}\,\cot{(2\theta_3)}= 
\hat{s}_2\, \cos{\delta}\,e^{-i\delta}\,,}
thereby confirming  our earlier identification of $\delta$ 
as the PMNS CP-violating
 phase. It follows that atmospheric neutrino oscillations must be  
less than maximal unless
the subdominant parameter $s_2$ vanishes.  (We show presently that
$\cos{\delta}$ cannot vanish in our model.)

We touch base with experiment with the relation:
\eqn\eh{\Delta_s= |m_{1}|^2-m_3^2 =
 m_3^2\tan^2{(2\theta_3)}\cos^2{\delta}.}
Using eqs.\ee,\ \eg\ and \eh, we  evaluate ---
 in terms of a single parameter --- three (as yet undetermined)
observables:   $M_{ee}$, the magnitude of the $e$-$e$ element
of ${\cal M}_\nu$ determining  the rate of neutrinoless double
beta-decay; $m(\nu_e)$,  the effective electron-neutrino mass (which in our
case is  simply $|m_1|$), and the sum $\Sigma$ of the magnitudes of the three
neutrino masses (such as has been  constrained astrophysically). We
find:
\eqn\ej{M_{ee}= {\sqrt{\Delta_a}\over y}, \quad\quad
m(\nu_e)=M_{ee}\sqrt{1+y^2}, \quad\quad 
\Sigma=M_{ee} + 2m(\nu_e)\,,}
where $y\equiv\tan{(2\theta_3)}\cos{\delta}$. 
 Any constraint or determination of one of these 
 observables constrains or determines the other two. 
Furthermore, from the experimental result
 $\tan{(2\theta_3)}\simeq 2.5$ we find that any such measurement constrains or
determines  the CP-violating phase $\delta$. For  any of the three
observables to be  large enough to be measured, $|\delta|$ must be large
(but less than  $\pi/2$ where $y=0$ and  neutrino masses diverge).
Additionally, we have $y=|s_2\tan{(2\theta_1)}|$ from eq.\esnd.
We exhibit these results for several choices of $\delta$ in the table below:

\medskip\medskip

\indent \vbox{\offinterlineskip
\hrule
\halign{&\vrule#&
   \strut\quad\hfil#\quad\cr %#&\quad#&\quad#\quad\cr
height2pt&\omit&&\omit&&\omit&&\omit&&\omit&&\omit&\cr
&$\delta=$\hfil&&$0^\circ$&&$45^\circ$&&$60^\circ$&&$
   86^\circ$&&$88^\circ$&\cr
height2pt&\omit&&\omit&&\omit&&\omit&&\omit&&\omit&\cr
\noalign{\hrule}
height2pt&\omit&&\omit&&\omit&&\omit&&\omit&&\omit&\cr
&$M_{ee}=$\hfil&&20&&28&&40&&292&&574&\cr
&$m(\nu_e)=$\hfil&&54&&57&&64&&297&&576&\cr
&$\Sigma=$\hfil&&128&&142&&168&&1186&&1726&\cr
&$|\sin{\theta_2}\tan{(2\theta_1)}|=$\hfil&&2.50&&1.77&&1.25&&
   0.17&&0.09&\cr
height2pt&\omit&&\omit&&\omit&&\omit&&\omit&&\omit&\cr}
\hrule}

\medskip\medskip

\noindent All  masses are given in milli-electron-volts. The last row
is not valid for
the special case $s_2=\cot{(2\theta_1)}=0$. The choice
$\delta=60^\circ$ yields a value for
$\Sigma$  that is barely compatible with the most severe
cosmological upper bound\rd. In this context, 
a value for $M_{ee}$  compatible with an alleged
observation\re\  of neutrinoless double beta decay or a value of $m(\nu_e)$ 
detectable  at Katrin\rg\  would require $\Sigma$ to exceed several
alleged cosmological bounds.

\medskip

We conclude with  a simple model  yielding  our ansatz \eb.
It involves three Higgs doublets
($h_e,\; h_\mu,\; h_\tau$) whose vevs and  Yukawa
couplings generate flavor-diagonal charged lepton masses
and   also  provide
Dirac masses for  the doublet neutrinos which, in conjunction with
flavor-conserving Majorana masses of the heavy singlet neutrinos,
yield see-saw neutrino masses with the texture of our ansatz.
 Assignments of our flavor quantum number 
to the various Higgs and left-handed lepton  fields are
given in the table below:
$$\vbox{\centerline{ The Flavor Quantum Number}\medskip
\settabs 6 \columns
\+&$D_e$ \  \ \  0&$e^+$ \ \ \ 0&$N_e$ \  \  \ 0\ &$h_e$\ \  \ \ 0&\cr
\+&$D_\mu$ \ +1&$\mu^+\,$ \ --1&$N_\mu$ \ --1/2&$h_\mu\,$ \ --1/2&\cr
\+&$D_\tau\,$ \ --1&$\tau^+$ \ +1&$N_\tau\,$ \ +1/2&$h_\tau$ \ +1/2
&\cr}$$
where the $D_\l$ and $h_\l$ are weak doublets.
The vev of $h_e$ generates arbitrary flavor-diagonal
masses of the three charged leptons (and could be responsible for
quark masses as well). 

Flavor-conserving bare mass terms for the heavy singlet states
$N_\l$ yield a heavy singlet mass matrix $M$ of the form:   
\eqn\mmt{ M=\pmatrix{F&0&0\cr 0&0&G\cr 0&G&0\cr}\,,}
whilst flavor-conserving  Yukawa couplings of singlet to doublet neutrino
states   yield
a  Dirac neutrino mass matrix $m$ of the form:
\eqn\ymm{m= \pmatrix{a&b&c\cr 0&d&0\cr 0&0&e\cr}\,.}
The   first column  of eq.\ymm\ arises  from $\langle h_e\rangle$,
the second from $\langle h_\mu\rangle$ and the third from 
$\langle h_\tau\rangle$.

\vfill\eject
Eqs.\mmt\ and \ymm\  enable the see-saw mechanism, yielding
the  light neutrino mass matrix:
\eqn\lnm{{\cal M}_\nu = m\,M^{-1}\,m^{TR}=
\pmatrix{(a^2/F+2bc/G)&cd/G&be/G\cr cd/G&0&de/G\cr
be/G&de/G&0\cr}\,,}
which is of the form \eb, as promised. 

\bigskip
\centerline{\bf Acknowledgment}\medskip

Critical comments of Andrew G. Cohen are much appreciated.
This research has been supported in part by the Department
of Energy under grant number DE-FG-02-01ER-40676.

\listrefs
%\figures
%\fig{1}{The $\delta$ dependence of $M_{ee}$, $m{\nu_e}$ (dotted)  and 
%$\Sigma$ (dashed) with $\tan{(2\theta_3)}=2.5$}

\parindent=20pt
\bye

 two well
established priors. We know that 
the subdominant mixing angle is small: $\sin^2{\theta_2 }<0.05$, and
that $\vert\Delta_a\vert \simeq 30\,\vert\Delta_s\vert$. Thus, in our
analysis we hereafter  work to first order in $\sin{\theta_2}$ and set
$|m_1|=|m_2|$. Errors due to these approximations are readily
estimable and small.

\bye

\bye